# Investigation of the anisotropy of dissipation processes in single crystal of Yba$_2$Cu$_3$O$_{7-\delta}$ system

J.G.Chigvinadze**, A.A. Iashvili**, J.V. Acrivos*, T.V. Machaidze**, T.Wolf***
*San Jose' State University, San Jose' CA 95192-0101, USA
**E. Andronikashvili Institute of Physics, 0177 Tbilisi, Georgia
*** Forschungszentrum Karlsruhe Institut für Festkörperphysik, Germany 76021 Karlsruhe, Germany

*By means of contactless mechanical method of the measurement of energy losses in superconductors, the anisotropy of dissipation processes has been studied in single crystal high-temperature layered superconductors of YBa$_2$Cu$_3$O$_{7-\delta}$ system, being in mixed state. The observed anisotropy of energy losses indicates the possibility of the existence of the symmetry of order parameter of d$_{x2-y2}$ type in these single crystals.*

## INTRODUCTION

The first direct experimental evidence of d$_{x2-y2}$ symmetry of order parameter (SOP) in high-temperature superconductors (HTSC) was given early [1], but after the existence of superconductivity was shown possible within the frames of the Hubbard model caused by the pairing of electrons in the exchange of fluctuation anti-ferromagnetic excitations with the parity of orbital angular momentum L=2 symmetry [2-5]. Though a fluctuation mechanism could not initially explain the high values of the observed transition temperature for superconductivity, T$_c$ [6] the phenomenology and self-consistent model used in the NMR data [7], showed that a pairing mechanism with a constant interaction may lead to an order parameter of d$_{x2-y2}$ symmetry with the values observed for T$_c$. These predictions for the existence of d$_{x2-y2}$ symmetry and the first direct experiments, confirming the existence of such symmetry arose a great interest in carrying out the experiments [8] for determination of the symmetry of order parameter in HSTC.

The experimental method for determining the SOP is a macroscopic one, based on the use of the dependence of tunnel current sensitivity in a Josephson contact on the phase of wave function of Cooper pair, which for a d-wave SOP depends on the direction of **k** vector, though such a conclusion was not immediately evident. In 1968 A.H. Silver and J.E. Zimmerman [9] observed, for the first time, the half-value of quantum magnetic flow $\phi_0$, ($\phi_0$/2) in Josephson contact. Later, G.E. Volovik and V.P. Mineev [10] proposed for the first time, that $\phi_0$/2 flow is connected with zeros of superfluid order parameter. Then B. Geshkenbein, A. Larkin and A. Barone [11] first noted that the tunnel current becomes zero, when the normal to the surface is oriented along the direction, in which the slit in the excitation spectrum is set to zero. Finally, M. Sigrist and T. M. Rice in 1992 [12] suggested to use the sensitivity of Josephson effect to the jump of the SOP phase for testing d$_{x2-y2}$ symmetry in HTSC. Immediately after this, the direct experiments were carried out to observe the d-wave SOP in the experiments of three types by using Jepheson contacts. One type of experiments measured the interference effects [1-13], the second type measured the modulation of the critical current by magnetic fields [14] and the third type used the method of tri-crystal magnetometry of the rings of super-conducting epitaxial films, allowing to observe the effect quantization of the magnetic flow with $\phi_0$/2 in the *YBa$_2$Cu$_3$O$_{7-\delta}$ system (*YBaCuO, GdBaCuO) Tl2201 and Bi2212 films [15]. However, it should be noted that even now experiments are still carried out to investigate the anisotropy of the super-conducting properties in HTSC, to ascertain whether the cause for the d$_{x2-y2}$ SOP is determined by individual Cu atomic orbitals or the macroscopic lattice effects [16]. Hartree Fock SCF calculations of molecular orbitals also show that the highest occupied molecular orbitals (HOMO) for Cu$_4$O$_4$ extended states with continuous overlap of electron states along the a/b diagonal direction in the CuO$_2$ plane can form Bose-Einstein pairs and that the O atoms are important in the superconducting wave function [16b,c].

## EXPERIMENTAL

Among the methods of the observation of the d-wave SOP in HTSC, the static mechanical method of measuring the magnetic moment, gave the first macroscopic evidence for the possibility of of d$_{x2-y2}$ symmetry, and for the measurements connected with this anisotropy of internal pinning [17]. In the present work we use the direct dynamic method of measuring the energy losses, proposed earlier [18]. This method gives the possibility to investigate not only the anisotropy of internal pinning, but the anisotropy of dissipated energy in HTSC, caused by internal pinning. The work deals just with the study of this anisotropy.

The method uses the oscillations of super-conducting sample, suspended from the thin elastic thread in the magnetic field, which is perpendicular to its surface and creates the Abrikosov vortex lattice (AVL) [19]. The oscillation frequency of the sample is changed when the vortex lines penetrate the sample. The action of the magnetic field on these vortices makes contribution to the torque of suspended system [20,21]. At small oscillation amplitude, the main mass of vortices is fixed on pinning centers, thereby causing the change of the oscillation frequency. As the oscillation amplitude increases, the most weakly fixed vortices begin to separate from the pinning centers, leading to an additional dissipation of energy, connected with their following motion in the volume of superconductor.

The theoretical analysis of our experiments in 1973 for superconductors of the II type [20, 21], allowed V.P. Galaiko to predict that even at small values of oscillation amplitudes there is a finite concentration of free, $C_{fr}$ and locked, $C_{lc}$ vortices, the interaction between which causes energy dissipation [22]. Further, these experiments were carried out on ceramic samples of high-temperature superconductors and the elastic-viscous properties of Abrikosov vortex structure were studied [23-24]. When a single crystal HTSC sample is suspended from a thin and elastic thread, with the c-axis parallel to the oscillation axis of the suspended system and the external magnetic field, $H_0 > H_{c1}$ is directed normal to c-axis in basal plane ab, the energy dissipation can be measured as a function of angle of rotation $\theta$ in basal plane. The anisotropy of the dissipation energy measured is determined by the internal pinning of high-temperature superconductor. We describe experiments carried out on $YBa_2Cu_3O_{7-\delta}$ single crystals of $2.1 \times 1.75 \times 0.2 mm^3$ dimensions, fabricated in the laboratory of Thomas Wolf in Karlsruhe (Germany). The apparatus for measuring the anisotropy of the dissipation energy processes in a single-crystal HTSC (FIG. 1) consists of **1:** a crystal holder that places its c-axis parallel to the axis of **2:** an aligned glass rod which is a part of **3:** a light disc made of duraluminum on its upper part with a moment of inertia $I = 2.255\ gr.cm^2$. Rod 2 is suspended from **4:** a thin elastic thread made of phosphor bronze (radius=50µm, length=170mm) that makes axial-torsional oscillations at time intervals registered by **5:** a mirror reflecting light on: **6:** two photo-sensors placed at a fixed distance from each other [25]. The assembly 1 to 5 was contained in a Dewar in at T= 90K. The logarithmic decrement of oscillation damping $\delta_{ab}$ is measured by the time interval from the oscillating mirror 5 reflections, between the photo sensors, located in the center, 6 was used for measuring the oscillation frequency of suspended system. By means of electronic unit the information from these sensors entered the computer, which processed the information and controlled the experiment. The magnetic field perpendicular to oscillating system and to c-axis of the sample was created by **7:** an electromagnet which could be rotated about the vertical axis allowing us to study the anisotropy of dissipation processes in the basal plane of single crystal HTSC. A small uniform magnetic field could created in any direction by **8, 9, 10:** additional Helmholtz coils, placed in the lower part of suspended system.

## RESULTS AND DISCUSSION

The measurement of the anisotropy of the dissipation processes (FIG. 2, 3) connected with internal pinning in single crystal high-temperature $YBa_2Cu_3O_{7-\delta}$ superconductor is described as follows:

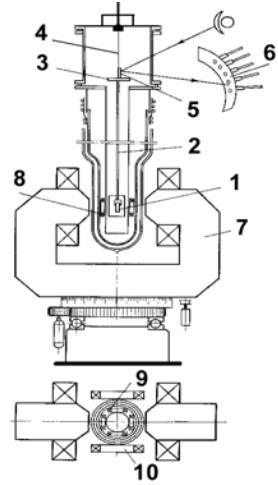

**FIG. 1:** Schematic of the apparatus used for the investigation of dissipation processes in superconductors: **1:** crystal holder, **2:** aligned glass rod, **3:** disc, **4:** thread made of phosphor bronze, **5:** mirror, **6:** photo sensors, **7:** electromagnet, **8, 9, 10:** Helmholtz coils.

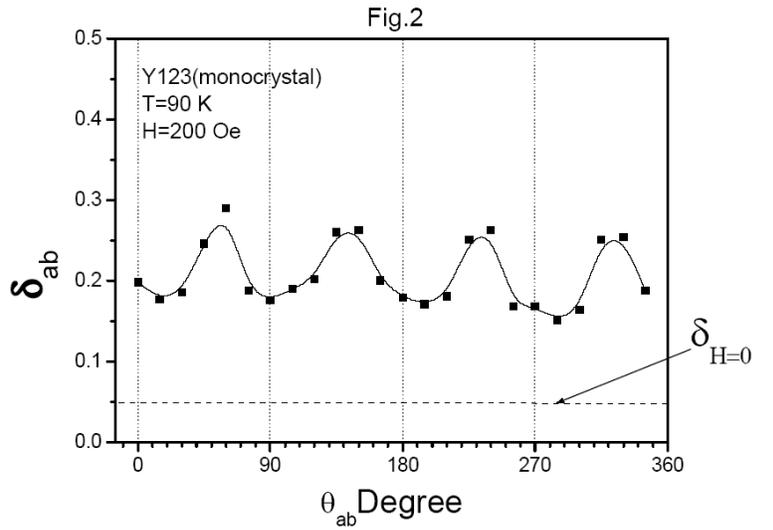

**FIG. 2:** Dependence of logarithmic decrement of oscillation damping $\delta$ on $\theta_{ab}$ angle in the basal plane of the single crystal.

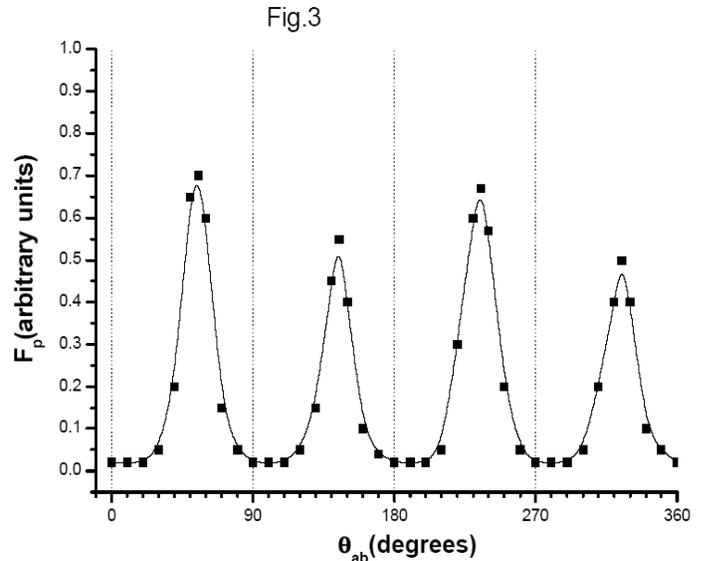

**FIG. 3:** Dependence of the pinning force $F_p$ on the $\theta_{ab}$ angle in the basal plane of single crystal.

The four minima in $\delta_{ab}$ in the basal plane: $\pi/2$, $\pi$, $3\pi/2$ and $2\pi$ (FIG. 2) indicate that when vortices are oriented along these directions, the energy dissipation is maximum when measurements are carried out at H=200Oe, T=90 K.

Since the anisotropy of energy dissipation in the basal plane is caused by internal pinning, it is important to also determine the nature of the dependence of the internal pinning force on the direction of external magnetic field with respect to **a** and **b** axes in the basal plane. Due to the low magnitude of pinning force in the small fields, the experiment was carried out at H=4200Oe, T=90K. In order to measure the pinning force, we use a specially developed mechanical (currentless) method [26] for direct measurement of pinning force in superconductors. The method is based on the measurements of counter moments of the forces acting on the superconducting sample from the side of magnetic field and torsion head. We measure the dependence of the angle of rotation of sample $\varphi_2$ on the angle of rotation of torsion head $\varphi_1$, rotation of which is passed to the super-conducting sample by the thread of suspension. In order to avoid the phenomena of inertia, the torsion head was rotated with low constant velocity $\omega=4.7*10^{-3}$ rad.sec$^{-1}$. According to the dependence of $\varphi_2$ on $\varphi_1$, one can determine the torque $\tau$ acting on super-conducting sample, which appears as the deviation of vortex ensemble from the direction of external magnetic field and at the rotation of upper head $\tau \sim K(\varphi_1-\varphi_2)$, where K is the stiffness factor of the suspended system. By measuring the dependence of the torque on the angle of rotation $\theta_{ab}$ between the direction of external magnetic field with respect to **a** and **b** axes of the basal plane, one can obtain the information on the character of the anisotropy of internal pinning in **ab** plane. The experiments (FIG. 3) show the dependence of internal pinning force on the angle between the external magnetic field H and the direction of basal axes in standard units, and clearly distinguish directions $\pi/2$, $\pi$, $3\pi/2$ and $2\pi$, corresponding to the orientations of vortices along **a** and **b** axes. In these directions the internal pinning has maximum. The rectangular super-conducting single crystal is not of disc shape, the lengths of vortices in different directions are not equal. $F_p$ is greater along one direction, $\theta_{ab}=\pi/4$ over $\theta_{ab}=3\pi/4$ by ~ 20 % which may be related to the sample dimensions, is much less than the maximum difference in pinning force versus $\theta_{ab}$ (FIG. 3).

## CONCLUSION

The characteristic curves of energy dissipation (FIG. 2) and pinning forces (FIG. 3) are consistent with $d_{x2-y2}$ SOP.

## ACKNOWLEDGEMENT
Work was supported by ISTC G-389 grant.